\documentclass[prb,aps,twocolumn,showpacs]{revtex4}

\newcommand{\xn}{y_{n}}
\newcommand{\yn}{z_{n}}
\newcommand{\nv}{\mathbf{n}\alpha}
\newcommand{\xnn}{u_{\nv}}
\newcommand{\un}{v_{n}}
\newcommand{\vn}{w_{n}}
\newcommand{\unn}{v_{\nv}}
\newcommand{\vnn}{w_{\nv}}
\newcommand{\unf}{v_{n+1}}
\newcommand{\vnf}{w_{n+1}}
\newcommand{\unb}{v_{n-1}}
\newcommand{\vnb}{w_{n-1}}
\newcommand{\dxn}{\dot{y}_{n}}
\newcommand{\dyn}{\dot{z}_{n}}
\newcommand{\dun}{\dot{v}_{n}}
\newcommand{\dvn}{\dot{w}_{n}}
\newcommand{\dln}{\delta l_{n}}
\newcommand{\dcn}{\delta \chi_{n}}
\newcommand{\Kb}{K_{b}}
\newcommand{\cof}{\cos(\varphi/2)}
\newcommand{\sif}{\sin(\varphi/2)}

\newcommand{\Ui}{U_{\mathrm{i}}}
\newcommand{\Uo}{U_{\mathrm{o}}}
\newcommand{\ijk}{\langle ijk \rangle}
\newcommand{\jkl}{\langle jkl \rangle}
\newcommand{\ljk}{\langle ljk \rangle}
\newcommand{\ij}{\langle ij \rangle}
\newcommand{\jk}{\langle jk \rangle}
\newcommand{\Kone}{K_{1}}
\newcommand{\Ktwo}{K_{2}}
\newcommand{\Kthr}{K_{3}}
\newcommand{\Kfou}{K_{4}}
\newcommand{\Lij}{\delta l_{ij}}
\newcommand{\Ljk}{\delta l_{jk}}
\newcommand{\Lkl}{\delta l_{kl}}
\newcommand{\dfi}{\delta \varphi_{ijk}}
\newcommand{\dfio}{\delta \varphi_{jkl}}
\newcommand{\ijkl}{\langle ijkl \rangle}
\newcommand{\il}{\{il\}}
\newcommand{\Kfiv}{K_{5}^{m}}
\newcommand{\Ksix}{K_{6}^{m}}
\newcommand{\Ksev}{K_{7}^{m}}
\newcommand{\Keig}{K_{8}^{m}}
\newcommand{\dxi}{\delta \chi_{ijkl}}

\newcommand{\ux}{u_{x}}
\newcommand{\uy}{u_{y}}

\newcommand{\vx}{v_{x}}
\newcommand{\vy}{v_{y}}

\newcommand{\wxx}{w_{xx}}
\newcommand{\wyy}{w_{yy}}
\newcommand{\wy}{w_{y}}
\newcommand{\wxy}{w_{xy}}
\newcommand{\kx}{k_{x}}
\newcommand{\ky}{k_{y}}
\newcommand{\kxt}{\kx^{2}}
\newcommand{\kyt}{\ky^{2}}
\newcommand{\omb}{\omega_{b}}
\newcommand{\ombt}{\omb^{2}}

\newcommand{\dxy}{dx \, dy}

\newcommand{\Kxy}{C_{1}}
\newcommand{\Kxyp}{C_{2}}
\newcommand{\Kz}{D_{1}}
\newcommand{\Kzp}{D_{2}}
\newcommand{\tKxy}{\tilde{\Kxy}}
\newcommand{\tKxyp}{\tilde{\Kxyp}}

\newcommand{\sq}{\sqrt{3}}
\newcommand{\CC}{\mathbf{C}_{h}}
\newcommand{\TT}{\mathbf{T}}
\newcommand{\TTparl}{T_{\|}}
\newcommand{\TTD}{\mathbf{T}_{D}}

\newcommand{\Ri}{\mathbf{R}_{i}}
\newcommand{\Rij}{\mathbf{R}_{ij}}
\newcommand{\Ch}{C_{h}}
\newcommand{\ko}{\mathbf{k}}
\newcommand{\kperp}{\mathbf{k}_{\bot}}
\newcommand{\kparl}{\mathbf{k}_{\|}}
\newcommand{\kp}{k_{\|}}
\newcommand{\kpe}{k_{\bot}}
\newcommand{\aper}{a_{\|}}
\newcommand{\dc}{d_{c}}

\newcommand{\aone}{\mathbf{a}_{1}}
\newcommand{\atwo}{\mathbf{a}_{2}}
\newcommand{\Rnm}{\mathbf{R}_{nm}}
\newcommand{\Na}{N_{a}}
\newcommand{\No}{N_{1}}
\newcommand{\Mo}{M_{1}}
\newcommand{\kt}{\mathbf{k}_{1}}
\newcommand{\GG}{\mathbf{G}}
\newcommand{\bone}{\mathbf{b}_{1}}
\newcommand{\btwo}{\mathbf{b}_{2}}
\newcommand{\dl}{\delta l}
\newcommand{\no}{n_{1}}
\newcommand{\nt}{n_{2}}
\newcommand{\rKo}{\mathbf{K}_{1}}
\newcommand{\rKt}{\mathbf{K}_{2}}

\usepackage{graphicx}
\usepackage{amsmath}
\begin{document}

\title{Simple empirical model for vibrational spectra of single-wall
carbon nanotubes}

\author{Yu.N. Gartstein}
\affiliation{Department of Physics, The University of Texas at
Dallas, P. O. Box 830688, FO23, Richardson, Texas 75083}
%\date{\today}
\begin{abstract}
A simple empirical model and approach are introduced for
calculation of the vibrational spectra of arbitrary single wall
carbon nanotubes. Differently from the frequently used force
constants description, the model employs only invariant quantities
such as variations of lengths and angles. All the salient
qualitative features of vibrational spectra of nanotubes naturally
follow from the vibrational Hamiltonian of graphene upon its
isometric mapping onto a cylindrical surface and without any
\textit{ad hoc} corrections. A qualitative difference with
previous results is found in a parabolic, rather than a linear,
long wavelength dispersion of the transverse acoustic modes of the
nanotubes. The parabolic dispersion is confirmed and elucidated in
the provided continuum analysis of the vibrations. We also discuss
and use an alternative definition of the nanotube unit cell with
only two carbons per cell that illustrates a ``true'' longitudinal
periodicity of the nanotubes, and of the corresponding Brillouin
zone.
\end{abstract}


\maketitle

\section{Introduction}

Vibrational spectra of individual single wall carbon nanotubes are
of considerable interest and have been calculated previously
within different frameworks such as an empirical force constant
model,\cite{book1,phononreview,cao} \textit{ab initio}
studies\cite{abini1} and tight-binding molecular
dynamics.\cite{tbmd} It is known that the higher-frequency part of
the  nanotube spectra is relatively well represented already by
the zone folding of the graphene spectrum. The lower-frequency
part, however, has generic features owing to the one-dimensional
character of nanotubes. Particularly, the spectra exhibit four
types of acoustic modes with vanishing frequencies: one
longitudinal, two transverse and one twisting. The existence of
these modes has to do with general considerations -- displacements
of the tube as a whole along and perpendicular to its axis, and
the rotation of the tube about the axis do not cost energy --
rather than with specific nanotube interactions. Analogous
vibrations were also discussed in the context of quantum wires  as
dilatational, flexural and torsional modes.\cite{nishi1} These
modes are important contributors to the low-temperature quantized
thermal conductance of such phonon
waveguides.\cite{rego1,expquant}

The frequently used force constant model of
Ref.~\onlinecite{book1} was first developed for planar graphene
based on the experimental data for graphite and then adapted for
nanotube geometries. Direct application of the graphene force
constant values was not found to lead to zero frequencies
 for all four modes mentioned above. To overcome
this difficulty, special curvature corrections were introduced to
the force constants.\cite{book1} In this paper we develop another
empirical model, where the harmonic vibrational Hamiltonian of
graphene is built using only ``invariant'' quantities such as
variations of bond lengths, interbond and dihedral angles. Such a
description is similar in spirit to used in conformational
analysis and stereochemistry (see, e.g.,
Ref.~\onlinecite{stereo1}) and in bond models for vibrations in
covalent semiconductors (Ref.~\onlinecite{cardona} and references
therein).  Further mapping of the graphene Hamiltonian onto a
cylindrical surface of the nanotubes then allows to derive all
features of arbitrary nanotube spectra ``naturally'', without any
curvature corrections. This way the idea of
Ref.~\onlinecite{book1} of using the same type of vibrational
Hamiltonian for both graphene and nanotubes turns out to be
realized with no need for \textit{ad hoc} modifications.

We find a qualitative difference with the previously published
results  in a parabolic, rather than a linear, dispersion of the
transverse acoustic modes of nanotubes. This parabolic dispersion
is further illustrated in the corresponding continuum model of
vibrations and is in agreement with the analysis of vibrations of
elastic cylindrical shells.\cite{bookmarkus} The continuum model
also shows the origin of another salient long wavelength feature
of the nanotube spectra: a coupling of the longitudinal acoustic
with the breathing mode. We believe the parabolic character of the
low-frequency part of the transverse mode dispersion is quite
generic similarly to the well-known bending waves of
rods\cite{ll8} with wavelengths much longer than the rod size.
Such a parabolic dispersion was, e.g., calculated for the lowest
flexural modes of quantum wires;\cite{nishi1,rego1} recent
applications of elastic cylinder models also include vibrations of
cytoskeletal filaments and
microtubules.\cite{sirenko1,sirenko2,bookphonons}
 To our knowledge, the
parabolic dispersion was not calculated previously for carbon
nanotubes.

With our approach we can easily calculate vibrational spectra of
arbitrary $(N,M)$ nanotubes. In doing this, we also employ an
alternative definition of the nanotube unit cell with only two
carbons per cell as opposed to possibly many carbon atoms of the
conventional definition.\cite{book1} This way a ``true''
longitudinal periodicity of the nanotubes is elucidated. The
period is a projection of one of the primitive vectors onto the
nanotube axis, the primitive vector itself being in general not
parallel to the axis. The  Brillouin zones can correspondingly be
wider than usually used and the total number of the vibrational
branches turns out to be just $6\dc$, where $\dc$ is the greatest
common divisor of $N$ and $M$.

For numerical computations, we will be using some parameterization
of the model, specifically based on a set of data from
Ref.~\onlinecite{book1} as well as on some experimental data.
However, the numerical computations here serve mostly illustrative
and qualitative purposes.

\section{\label{VH}Vibrational Hamiltonian}

Despite the fact that atoms of carbon nanotubes are arranged in a
3-$d$ fashion, the excitations of nanotubes can be described in
the way very similar to excitations of the planar graphene. To
clearly see this connection for vibrational excitations, one can
use local (position-dependent) coordinate systems for atomic
displacements and the elastic potential energy written in an
``invariant'' form. This way the qualitative transformation from
vibrational spectra of an infinite plane to spectra of curved
cylindrical structures appear naturally without \textit{ad hoc}
corrections.

Following Ref.~\onlinecite{book1}, we also consider carbon-carbon
elastic interactions up to the fourth nearest neighbor. The
harmonic potential energy $U=\Ui+\Uo$ is however expressed as a
function of only invariant quantities such as variations of bond
lengths and various angles. The first term $\Ui$ would correspond
here to in-plane deformations of graphene and in general requires
ten elastic parameters $K_{i}^{m}$:
\begin{subequations}\label{Ui}
\begin{eqnarray}
\Ui & = & \sum_{\ijk}\left[\Kone(\Lij^{2}+\Ljk^{2}) + \Ktwo
\dfi^{2}
+ \Kthr\Lij\Ljk \right. \nonumber \\
& + & \left. \Kfou\dfi(\Lij+\Ljk)
 \right] \label{Uia} \\
& + & \sum_{m=3}^{4}\sum_{\ijkl}^{\il=m} \left[\Kfiv\Lij\Lkl +
\Ksix\dfi\dfio \right. \nonumber \\
& + & \left. \Ksev(\dfi\Lkl+\dfio\Lij) \right], \label{Uib}
\end{eqnarray}
\end{subequations}
while term $\Uo$ would describe out-of-plane graphene distortions
and needs three elastic parameters:
\begin{equation}\label{Uo}
\Uo=\sum_{m=2}^{4}\sum_{\ijkl}^{\il=m}\Keig \dxi^{2}.
\end{equation}
\begin{figure}
\includegraphics[scale=0.5]{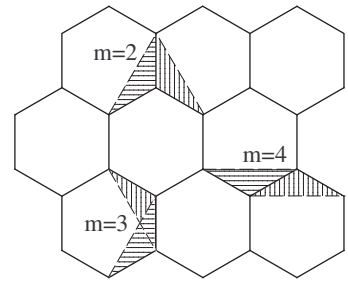}
\caption{\label{elint1}Nomenclature of possible triangular
plaquette pairs adjacent along a nearest-neighbor carbon-carbon
bond. To guide the eye, individual adjacent plaquettes are filled
with lines at different angles.}
\end{figure}

The structure of Eqs.~(\ref{Ui},\ref{Uo}) can be conveniently
thought of in terms of triangular plaquettes $\ijk$ formed by
bonds $\ij$ and $\jk$ connecting nearest carbons $i$ and $j$, and
$j$ and $k$, respectively. Variation of bond $\ij$ length is
denoted $\Lij$, and variation of the inter-bond angle at the
common carbon $j$ denoted $\dfi$. Correspondingly, Eq.~(\ref{Uia})
completely describes the deformation energy of individual
plaquettes. Equations (\ref{Uib},\ref{Uo}), on the other hand,
completely describe the interactions of deformations on
neighboring plaquettes; specifically, $\ijkl$ stands for
plaquettes $\ijk$ and $\jkl$ that are adjacent along bond $\jk$.
There are three different ways to form adjacent plaquette pairs
and notation $\il=m$ distinguishes them by indicating that carbons
$i$ and $l$ are the $m$th nearest neighbors ($m=2,3,4$), as
illustrated in Fig.~\ref{elint1}. (In case $m=2$, it is actually
plaquettes $\ijk$ and $\ljk$ that are adjacent along $\jk$. This
case is not explicitly included in Eq.~(\ref{Uib}) because of the
constraint that a sum of inter-bond angles for three plaquettes
surrounding a carbon atom is fixed.) The ``out-of-plane''
interaction, Eq.~(\ref{Uo}), involves the variations $\dxi$ of the
dihedral angles between corresponding plaquettes.

It is worth stressing that for the \textit{planar} graphene,
Eqs.~(\ref{Ui},\ref{Uo}) give the most general description of
harmonic interactions involving up to the forth nearest neighbors.
As such, it, of course, can reproduce a force constants
description. The latter would be derived by expanding
Eqs.~(\ref{Ui},\ref{Uo}) in the atomic pair differences.
Evidently, then there would be certain relationships between a
larger number of force constants as dictated by the invariance of
the potential energy with respect to overall
rotations.\cite{madelung} Specifically, ten elastic parameters of
Eq.~(\ref{Ui}) yield twelve force constants, comprising eight
in-plane constants of the type explicitly considered in
Ref.~\onlinecite{book1} and four constants mixing radial and
tangential displacements that were implicitly set to zero in that
reference. Three elastic parameters of Eq.~(\ref{Uo}) yield four
out-of-plane force constants.\cite{book1}

The advantage of using invariant quantities in
Eqs.~(\ref{Ui},\ref{Uo}) is that the same functional form of the
potential energy can be directly used when carbon atom positions
are (isometrically) mapped from the graphene plane onto the
cylindrical surface of a nanotube. When on the nanotube surface,
variations of the bond lengths and various angles just need to be
calculated from the carbon atom displacements using the actual
curved geometry. Of course, a new set of force constants
appropriate for the now curved geometry can be again derived
through the pair expansions. Force constants so obtained would
automatically obey the correct relationships to satisfy the
invariance with respect to rotations.

The standard translational invariance of the vibrational
Hamiltonian is preserved if displacements of carbons are expressed
not in terms of common $(xyz)$ coordinates but in terms of local
orthogonal coordinates $(uvw)$: $u$ - along the nanotube axis, $v$
- perpendicular to the tube axis and parallel to the tube surface,
and $w$ - perpendicular to the tube surface. These simple ideas
are illustrated in more detail in Appendix \ref{ringexample} for
an easier to follow example of the relationship between vibrations
of a linear chain of atoms and vibrations of a ring of atoms. For
the problem at hand, we use the local displacements
$(\xnn,\unn,\vnn)$ depending on carbon $\nv$ actual geometric
position on the cylindrical surface of the tube. (Here carbon
index $\nv$ consists of a 2-$d$ vector $\mathbf{n}$ specifying the
unit cell of the parent graphene plane and $\alpha=1,2$ specifying
one of the 2 carbons in the graphene unit cell.) With local
displacement bases in place, the invariance of the vibrational
Hamiltonian with respect to translations by graphene primitive
vectors is held on equal footing in graphene and nanotubes: wave
vectors $\ko$ will ``know'' only differences between neighboring
carbon indices $\nv$. The problem is thereby reduced to
calculation of the usual, ``graphene-like'', $6 \times 6$
dynamical matrix, but which naturally contains the correct mixing
of the ``in-plane'' and ``out-of-plane'' displacements in
nanotubes. Using proper quantization rules for the allowed phonon
wavevectors, one can then readily derive the vibrational spectra
of nanotubes of arbitrary chirality.

\section{\label{unitcells}Primitive cells and Brillouin zones}

The geometry of a single wall carbon nanotube is determined by the
chiral vector\cite{book1}
\begin{equation}\label{CC}
\CC= N\aone + M\atwo,
\end{equation}
where $\aone$ and $\atwo$ ($|\aone|=|\atwo|=a$) are two primitive
vectors of the 2-$d$ graphene (hexagonal) crystal structure.
Vector $\CC$ is perpendicular to the nanotube axis.
Conventionally,\cite{book1} the translational vector
\textit{parallel} to the tube axis is defined, which we denote
here as $\TTD$: $\TTD=t_{1}\aone+t_{2}\atwo$, where
$t_{1}=(2M+N)/d_{R}$, $t_{2}=-(2N+M)/d_{R}$ and $d_{R}$ is the
greatest common divisor of $(2N+M)$ and $(2M+N)$. The resulting
unit cell of the nanotube built of $\CC$ and $\TTD$ can contain
many carbons $\Na=4(N^{2}+M^{2}+NM)/d_{R}$, and the longitudinal
period $|\TTD|$ of chiral tubes be much larger than $a$. The
corresponding Brillouin zones (BZs) would then be narrow and
contain many excitation spectrum (whether vibrational or
electronic) branches.

In this paper we use an alternative picture of the unit cell and
BZ construction that is aimed at having as small number of
branches in the zone as possible. As described in more detail in
Appendix \ref{newbz}, this number of branches is determined by the
greatest common divisor of $N$ and $M$, denoted by $\dc$. The
total number of continuous branches in BZ is equal to $2\dc$ per
each degree of freedom of a carbon atom, that is, $6\dc$ for
vibrational excitations. This  corresponds to the nanotube unit
cell containing \textit{only two carbons} and which can, e.g., be
built with primitive vectors $\CC/\dc$ and $\TT$. Different from
vector $\TTD$, the translational vector $\TT=P\aone+Q\atwo$  is in
general \textit{not parallel} to the nanotube axis. As discussed
in Appendix \ref{newbz} (see Eq.~(\ref{PQ1})), integers $P$ and
$Q$ here satisfy condition $MP-NQ=\dc$. It is the projection of
vector $\TT$ onto the axis that determines the longitudinal period
\begin{equation}\label{aper1}
\aper=\sq \dc a^{2}/2\Ch
\end{equation}
and the width $2\pi/\aper$ of the BZ.

The number of branches in the BZ is related to the transverse
quantization of the 2-$d$ wave vector $\ko$ of the parent graphene
band excitations:
\begin{equation}\label{quant}
\ko \CC = 2\pi l,
\end{equation}
resulting in the appearance of one-dimensional sub-bands
characterized by the integer quantum number $l$. The construction
of the unit cell and BZ employed in this paper recognizes that
there would be only $\dc$ ``unique'' quantization levels (that is,
only $\dc$ independent integers $l$): all other allowed
$\ko$-vectors can be obtained with translations by graphene
reciprocal vectors. Many branches of the conventional\cite{book1}
BZ  would not exhibit gaps at that BZ boundary; they would
correspondingly become single continuous bands when properly
``unfolded'' in our construction. The quantization lines within
our BZs can span several hexagons of the graphene reciprocal
lattice (see example of Fig.~\ref{recipr1}).

\section{Vibrational spectra}

\begin{figure}
\includegraphics[scale=0.64]{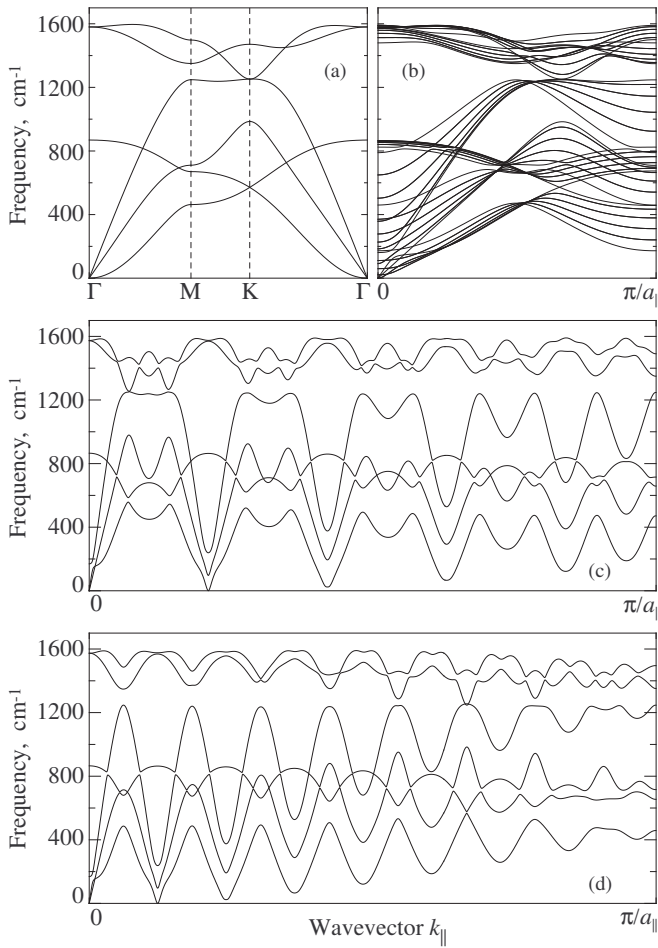}
\caption{\label{Spectra1}Model vibrational spectra: (a) Graphene;
(b) (10,10) tube; (c) (10,9) tube; (d) (16,1) tube. See text for
definition of $\aper$.}
\end{figure}

As was mentioned above, potential energy in
Eqs.~(\ref{Ui},\ref{Uo}) is capable of reproducing results of the
force constant model\cite{book1} for the graphene spectrum. Figure
\ref{Spectra1} (a), however, has been calculated with a
parameterization of elastic constants $K_{i}$ in
Eqs.~(\ref{Ui},\ref{Uo}) such as to achieve only a close
similarity to the published spectrum.\cite{book1} In our
calculations, we chose to slightly and somewhat arbitrarily modify
the tangential force constants $\phi^{(n)}_{t}$ from the
published\cite{book1} values so as to satisfy
$\phi^{(1)}_{t}+6\phi^{(2)}_{t}+4\phi^{(3)}_{t}+14\phi^{(4)}_{t}=0$.
The latter equality is required by the rotational invariance --
elastic energy should be zero for the overall rotation of the
graphene plane. Original constants\cite{book1} do not obey it. A
recently published new set of force constants\cite{newforce} also
does not satisfy this requirement. The overall scaling of elastic
constants was chosen here so as to reproduce graphene experimental
optical frequencies of 1580 and 868 cm$^{-1}$. Once the values of
elastic constants $K_{i}$ have been defined for graphene, the same
values are used to calculate spectra of nanotubes, examples of
which are shown in Figures \ref{Spectra1} (b)--(d). As discussed
in Sec.~\ref{VH}, we do not need the knowledge of force constants
of Ref.~\onlinecite{book1} for curved geometries because the
Hamiltonian used automatically preserves all invariance
requirements.

In displaying the nanotube spectra, we use the definition of BZ as
discussed in Sec.~\ref{unitcells} and Appendix \ref{newbz}.
 With that definition, the BZ contains only  $6\dc$
vibrational branches. For armchair tubes, $\aper=a/2$ and the BZ
in Figure \ref{Spectra1} (b) is twice as wide as the conventional
BZ, that picture would be restored by simple folding. Figure
\ref{Spectra1} (b) exhibits $\dc=10$ ``independent'' transverse
quantization levels. On the other hand, Figures \ref{Spectra1} (c)
and (d) with $\dc=1$ correspond to only one transverse
quantization level, that is, to six branches. Each polarization
band/branch exhibits a continuous evolution of the polarization
vectors with in general a strong $\kp$ dependence. It is worth
noting that transverse acoustic modes in this picture have their
frequency vanishing at finite $\kp$ (which, of course, lie in a
hexagon of the graphene reciprocal lattice other than the one
where $\kp=0$ is).

\begin{figure}
\includegraphics[scale=0.55]{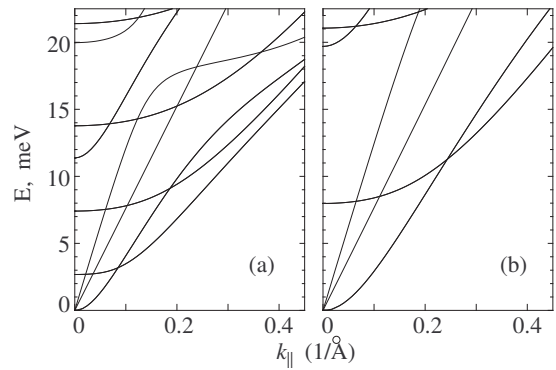}
\caption{\label{lowfreq}The low-frequency part of calculated
nanotube vibrational spectra: (a) (10,10) tube, (b) (10,0) tube.
Note that here wave vectors $\kp$ are measured with respect to
closest $\Gamma$ points of hexagons of the reciprocal graphene
lattice.}
\end{figure}

Figure \ref{lowfreq} shows low-frequency parts of the vibrational
spectra of the (10,10) and (10,0) tubes reduced to a single
hexagon of the reciprocal lattice. Panel (a) can be directly
compared to the published results\cite{phononreview} derived from
the model of Ref.~\onlinecite{book1}. Apart from the small
differences, such as values of acoustic velocities, likely caused
by our modification of the force constants, there is one
qualitative disparity. The dispersion of the transverse acoustic
modes in Figure \ref{lowfreq} is clearly seen to be parabolic in
contrast to a linear dispersion discussed in
Refs.~\onlinecite{book1,phononreview,tbmd}. The parabolic
dispersion is seen over a wider range of $\kp$ on going to tubes
of smaller radii, compare Figures \ref{lowfreq} (a) and (b). In
Sec.~\ref{continuum} we give an analytic confirmation of this
observation within a framework of a continuum mechanics.

\section{\label{continuum}Continuum analysis}

Analysis of deformations and vibrations of thin-walled elastic
cylinders goes back as far as to Rayleigh and Love; see, e.g.,
Refs.~\onlinecite{rayleigh, love, timoshenko, timoshenko1},
references therein, and Refs.~\onlinecite{bookmarkus,arnold} for a
dedicated analysis of vibrations.

In the case of a planar, graphene, structure, the continuum
potential elastic energy can be written as
\begin{eqnarray}\label{Uic}
\Ui & = & \frac{\rho}{2} \int \dxy \left[ \Kxy (\ux+\vy)^{2}
\right. \nonumber \\
& + & \left. \Kxyp  \left(
%(\ux-\vy)^{2} + (\uy+\vx)^{2}
(\uy+\vx)^{2}-4\ux\vy \right) \right],
\end{eqnarray}
\begin{eqnarray}\label{Uoc}
\Uo & = & \frac{\rho}{2} \int \dxy \left[ \Kz (\wxx+\wyy)^{2}
\right.
\nonumber \\
& + & \left. \Kzp (\wxy^{2}-\wxx\wyy) \right].
\end{eqnarray}
Here $x$ is a coordinate that would later become along the
cylinder axis and $y$ coordinate  along the cylinder
circumference,  $\rho$ is the mass density. Displacements fields
$u$, $v$ and $w$ would become, respectively, parallel to the
cylinder axis, parallel to its circumference, and perpendicular to
the cylindrical surface; $x$ and $y$ subindices denote the
differentiation over corresponding coordinates. This type of
deformation energy is well known in the continuum mechanics of
plates\cite{timoshenko,ll8} and can be readily derived from the
discrete form (\ref{Ui},\ref{Uo}). In the former picture, elastic
constants $\Kxy$, $\Kxyp$, $\Kz$ and $\Kzp$ in
Eqs.~(\ref{Uic},\ref{Uoc}) are expressed in terms of stretching
and bending rigidities and Poisson's ratio. In the latter
derivation, they would be expressed through constants $K$'s in
Eqs.~(\ref{Ui},\ref{Uo}). Note, however, that
Eqs.~(\ref{Ui},\ref{Uo}) should then be transformed only in the
context of purely acoustic deformations, in which two carbons in a
hexagon move ``in-phase''. That is why we will be dealing here
with the $3\times 3$ matrix in Eq.~(\ref{matrix}) rather than with
a $6\times 6$ dynamical matrix. One can develop a continuum model
that would include optical ``out-of-phase'' deformations as well.
The deformation energy of type of Eqs.~(\ref{Uic},\ref{Uoc}) was
already proved to be useful in studies of large deformations of
carbon nanotubes.\cite{yakobson}

\begin{figure}
\includegraphics[scale=0.55]{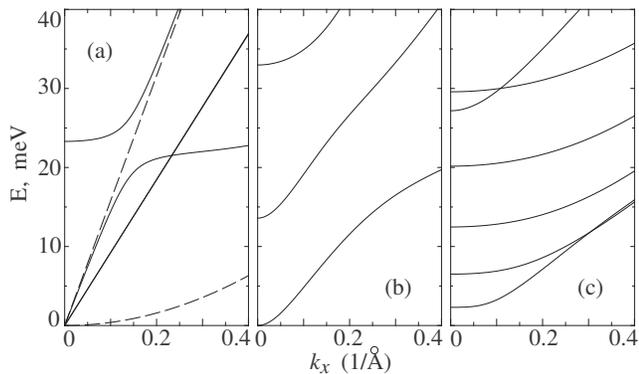}
\caption{\label{cont1}Branches of the vibrational spectra obtained
in the continuum model described in the text at (a) $\ky=0$, (b)
$\ky=1/R$, (c) $\ky=n/R$ with $n\geq 2$. Calculations were
performed for $R=6.78$ \AA \ that one would have for the (10,10)
tube. Elastic parameters used in calculations for this Figure
would correspond to graphene velocities $c_{la}=24$ km/s,
$c_{ta}=14$ km/s. Dispersion of the out-of-plane graphene
vibrations $\omega=\delta k^{2}$ was taken with $\delta=6\times
10^{-7}$ m$^{2}$/s (see Ref.~\onlinecite{abini1}), and
$\Kzp/\Kz=4\Kxyp/\Kxy$. Graphene results are shown by the dashed
lines for comparison. Dispersion of the graphene transverse
acoustic mode practically coincides with that of the nanotube
twisting mode.}
\end{figure}

The modification of Eqs.~(\ref{Uic},\ref{Uoc}) upon formation of a
cylindrical body in continuum mechanics is, e.g., discussed in
Refs.~\onlinecite{bookmarkus,timoshenko, timoshenko1}. As is also
shown in Appendix \ref{ringexample} for a discrete model, this
corresponds to a simple substitution in (\ref{Uic},\ref{Uoc}):
$\vy \rightarrow \vy + w/R$ and $\wy \rightarrow \wy -v/R$, $R$
being the cylinder radius. Using this substitution, one can easily
study the problem of small vibrations of a continuum cylinder. The
vibrational frequencies $\omega$ for the plane waves with a
two-dimensional wave vector $(\kx,\ky)$ are determined from the
eigenvalue equation
\begin{equation}\label{disp}
\omega^{2}\mathbf{d}=\mathbf{M}\mathbf{d},
\end{equation}
where displacement vector $\mathbf{d}=(u,v,w)$ and matrix
$\mathbf{M}(\kx,\ky)$ is
\begin{equation}\label{matrix}
\mathbf{M}\!\!=\!\!\left[ \!\!\!
\begin{array}{ccc}
\Kxy \kxt + \Kxyp \kyt & (\Kxy\!-\!\Kxyp)\kx\ky & i
A\kx/R \\
(\Kxy\!-\!\Kxyp)\kx\ky & \tKxy\kyt\! + \!\tKxyp\kxt & -iB\ky/R \\
-iA\kx/R & iB\ky/R & \ombt \!+\!\Kz(\kxt \!+\!\kyt)^{2}
\end{array}
\!\!\!\right].
\end{equation}
Here $\tKxy=\Kxy+ \Kz/R^{2}$, $\tKxyp=\Kxyp+\Kzp/R^{2}$,
$A=2\Kxyp-\Kxy$, $B= \Kxy + (\Kz+\Kzp/2) \kxt + \Kz \kyt$ and
$\ombt=\Kxy/R^{2}$.

In the case of the planar structure, $R\rightarrow\infty$,
Eqs.~(\ref{disp},\ref{matrix}) lead to two acoustic waves of
in-plane vibrations (with longitudinal  $c_{la}=\Kxy^{1/2}$ and
transverse $c_{ta}=\Kxyp^{1/2}$ velocities) and an acoustic wave
of the out-of-plane vibrations with a parabolic spectrum
$\omega=\Kz^{1/2}(\kxt+\kyt)$, see Figure \ref{cont1} (a).

For a cylinder/tube of a finite radius $R$, the transverse
quantization imposes a restriction on values of $\ky =n/R$, where
$n$ is an integer. Of special interest to us here are values of
$\ky$ equal to 0 and to $\pm 1/R$. The case of $\ky=0$
(Fig.~\ref{cont1} (a)) yields two of the four acoustic modes of
the tube with vanishing frequencies: (i) the longitudinal mode
with a low-frequency dispersion coinciding with the longitudinal
acoustic wave of the graphene and (ii) the twisting mode whose
dispersion is somewhat modified from the transverse acoustic mode
of graphene by virtue of the elastic constant $\Kzp$:
$c_{ta}=\tKxyp^{1/2}$. The longitudinal mode in the case of the
cylinder couples with the breathing mode, whose frequency at
$\kx=0$ is $\omb$ and whose dispersion is determined by constant
$\Kz$. As a result of this coupling, an anti-crossing behavior of
the branches arises as is clearly seen in the Figure.  (iii) The
two other, degenerate, acoustic modes of the tube spectrum,
usually referred to as transverse acoustic modes for carbon
nanotubes\cite{book1} or as flexural modes in other
applications,\cite{arnold,nishi1,sirenko1,sirenko2} correspond to
$\ky=\pm 1/R$ (Fig.~\ref{cont1} (b)). It is apparent from the
Figure
 that these modes have a parabolic spectrum. In fact,
one can easily show this analytically by the perturbation analysis
of (\ref{disp},\ref{matrix}) in $\kx$: contributions to the linear
coefficient in dispersion $\omega(\kx)$ exactly cancel. The
low-frequency parabolic dispersion of these modes, although
modified by the curvature rigidity, Eq.~(\ref{Uoc}), is mainly
determined by the in-plane stretching rigidity, Eq.~(\ref{Uic}).
Neglecting Eq.~(\ref{Uoc}), one would obtain the long wavelength,
for $\kx R \ll 1$, dispersion of this mode as
$\omega(\kx)=\left(2(\Kxy-\Kxyp)\Kxyp/\Kxy\right)^{1/2}\kxt R$.
Evidently, the parabolic character of the dispersion becomes even
more apparent for smaller-radius tubes. These modes correspond to
bending vibrations of the tube as a whole and, in this sense, are
similar to the bending vibrations of rods, whose generic
long-wavelength parabolic dispersion is well known.\cite{ll8}
Similarities between Figures \ref{cont1} and \ref{lowfreq} (a) are
obvious. Note, however, some quantitative differences caused by
different values of effective parameters.

\section{Summary and discussion}

We have described a simple empirical model, in which vibrations of
the graphene and individual single wall carbon nanotubes are
treated and calculated on the same footing. Differently from the
force constant model, our model uses only ``invariant''
quantities: variations of bond lengths, interbond and dihedral
angles. As a result, the isometric mapping from the planar
graphene onto a cylindrical surface of nanotubes automatically
preserves all the right relationships between equivalent force
constants. Importantly, all calculated vibrational spectra of
nanotubes correctly exhibit four types of acoustic excitations
with vanishing frequencies (one longitudinal, two transverse and
one twisting), which are derived naturally and follow from the
symmetries of the underlying system. Although the results obtained
are largely similar to the earlier published, we have also found
an important qualitative difference. The long-wavelength
dispersion of the transverse acoustic modes is shown to be
parabolic rather than linear. One consequence of this is that the
vibrational density of states should exhibit a one-dimensional
singularity near zero frequency. We cannot exclude that there can
be other physical implications of our finding, although this
apparently is not the case for the quantized ballistic thermal
conductance,\cite{rego1,expquant} for which the exact dispersion
law is irrelevant. These long-wavelength (wavelength much larger
than the tube radius) transverse modes correspond to bending
oscillations of a nanotube as a whole and, therefore, are similar
to bending vibrations of rods, whose low-frequency spectrum is
known to be parabolic.\cite{ll8} The parabolic dispersion is,
e.g., evident in calculations of the lowest flexural modes of
quantum wires.\cite{rego1,nishi1} The origin of the parabolic
dependence for nanotubes has been analytically illustrated  using
a continuum elastic model similar to used in the analysis of
vibrations of cylindrical shells.\cite{bookmarkus} The continuum
model also clarified the coupling between longitudinal acoustic
and breathing modes resulting in the anti-crossing behavior
clearly seen in the calculated spectra of nanotubes.

The simplicity of our model allows us to easily calculate
vibrational spectra of nanotubes of arbitrary chiralities. To
better handle such spectra, we employed an alternative definition
of the nanotube unit cell with only two carbons per cell. This
definition reveals a ``true'' longitudinal periodicity of carbon
nanotubes that can be substantially shorter than used in the
conventional definition.\cite{book1}

Being a result of  the straightforward isometric mapping, the
model presented in this paper, has its qualitative limitations.
For instance, it does not automatically yield the
curvature-induced softening of the phonon modes observed in
\textit{ab initio} studies such as that of the twisting
mode,\cite{abini1} or more complex dependencies for the breathing
mode.\cite{abini1,rbmmiklosh} Of course, these effects are not
deducible from the graphene properties. To describe them, the
model would need to include both the relaxation of nanotube
geometries (mapping would not be exactly isometric) as well as an
explicit dependence of the elastic energy on nanotube radius and
with the axial anisotropy.\footnote{See Appendix \ref{ringexample}
for a discussion of some of these issues for a simpler example.}
We discussed such a generalization in the context of uniform
deformations,\cite{gzb2} which, e.g., leads to the
chirality-dependent stiffness of nanotubes.\cite{mint1} We believe
the present model can also be further developed in this regard and
correspondingly parameterized by comparison with \textit{ab
initio} calculations. A more accurate parameterization is,
needless to say, required even for the present model. Such
quantitative aspects have not been pursued in this paper.

\begin{acknowledgements}
This work was started in the context of DARPA program on "Phonon
Engineering" funded under the contract No.~MDA 972-02-C-0044. I am
grateful to V.~Agranovich, R.~Baughman, A.~Efros, M.~Kertesz and
A.~Zakhidov for discussions and various inputs related to the
subject.
\end{acknowledgements}

\appendix
\section{\label{ringexample}A ring example}

\begin{figure}[b]
\includegraphics[scale=0.5]{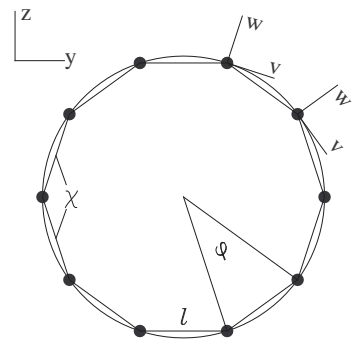}
\caption{\label{gring1}A ring of $N=10$ atoms connected by
interatomic ``bonds'' (thicker lines). The stretching rigidity
corresponds to variations of the bond lengths $l$, the bending
rigidity to variations of the inter-bond angles $\chi$. Number of
atoms $N$ (or the radius of the ring) determines the angle
$\varphi=2\pi/N$.}
\end{figure}

\begin{figure}[t]
\includegraphics[scale=0.5]{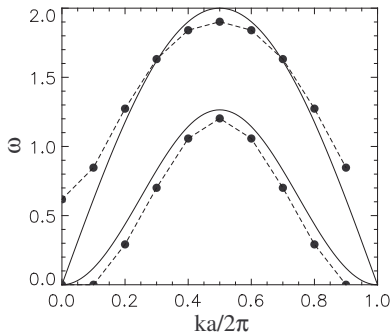}
\caption{\label{ring1}The vibrational spectra of the linear chain
(solid lines) and of the ring (filled circles) of atoms. The
parameters used for the plot are  $K/M=1$ and $\Kb/M=0.1$.}
\end{figure}

Here we study in more detail a simpler illustrative example of the
relationship of vibrations of an infinite linear atomic chain and
a ring of atoms. For clarity, displacements of atoms are
restricted to the plane in which a ring and chain belong.
Atom-atom interactions correspond to formation of bonds and result
both in stretching and bending rigidity of a linear chain. Figure
\ref{gring1} is a picture of the ring consisting of $N=10$ atoms,
which is assumed to preserve the nature of interactions in the
chain. The figure shows a common system of coordinates $(y,z)$ as
well as local systems $(v,w)$ related to the tangential and normal
displacements of the corresponding atom. Obviously, the kinetic
energy
\begin{equation}\label{kin}
T=\frac{M}{2}\sum_{n} (\dxn^{2} + \dyn^2) = \frac{M}{2}\sum_{n}
(\dun^{2} + \dvn^2),
\end{equation}
where $n$ is the 1-$d$ positional index of the atom along the ring
circumference. We first write the potential energy of a linear
chain in an invariant form as
\begin{equation}\label{u1}
U=\frac{K}{2}\sum_{n} \dln^{2} + \frac{\Kb l^{2}}{2}\sum_{n}
\dcn^{2},
\end{equation}
where the first term describes the stretching and second the
bending rigidity. Then we assume that the same functional form
holds for the ring. From the geometry of Figure \ref{gring1}, the
variation of the bond lengths, as a function of the local
displacements, is
\begin{equation}\label{l}
\dln=(\unf-\un)\cof + (\vnf+\vn)\sif,
\end{equation}
while the variation of the inter-bond angles
\begin{eqnarray}
l\cdot\dcn & = &(2\vn-\vnf-\vnb)\cof \nonumber \\
& + & (\unf-\unb)\sif. \label{c}
\end{eqnarray}

For the linear chain ($\varphi=0$ and $\un=\xn$, $\vn=\yn$),
Eq.~(\ref{l}) describes a conventional longitudinal stretching
while Eq.~(\ref{c}) would yield a conventional local curvature.
Tangential and normal vibrations in a linear chain are fully
decoupled. The corresponding two branches of the vibrational
spectrum are simply
\begin{equation}\label{chain}
\begin{array}{l}
\omega^{2}_{s}(k)  =  (4K/M)\sin^{2}(ka/2), \\
\omega^{2}_{b}(k)=(16\Kb/M)\sin^{4}(ka/2),
\end{array}
\end{equation}
where $a=l$ is the distance between the atoms along the chain.
This
spectrum is shown in Figure \ref{ring1} with solid lines.\\

 In the
curved system (our ring with $\varphi \neq 0$), on the other hand,
tangential and normal displacements \textit{are} coupled, as is
well known in the elasticity theory for curved
surfaces\cite{timoshenko,ll8} and clearly seen in
Eqs.~(\ref{l},\ref{c}). If we were to use the common system of
coordinates for displacements $(\xn,\yn)$, the atom contributions
to the potential energy, Eq.~(\ref{u1}), would be position
dependent and the translational invariance with respect to $n
\rightarrow n+1$ would be lost. With the local coordinates
$(\un,\vn)$, this invariance is preserved and one can directly use
the conventional transition to $k-$states, which would now be
quantized as
\begin{equation}\label{qua1}
k=(2\pi/Na) i, \ \ \ \ i=0, \ldots N-1.
\end{equation}
The unit cell length $a > l$ here is now the distance between the
atoms of the ring along its circumference. The derivation of the
spectrum from the equations of motion follows straightforwardly
from Eqs.~(\ref{kin}--\ref{c}), and the results are shown in
Figure \ref{ring1} with filled circles. In accordance with
Eq.~(\ref{qua1}), it is now a set of discrete frequencies. The
coupling between normal and tangential displacements resulted in
important qualitative modifications of the spectrum, which have
been obtained exactly and naturally. Particularly, one notices 3
zero-frequency modes. Two of them (with $i=1$ and $i=N-1$, or
$i=-1$, in Eq.~(\ref{qua1}) -- precisely one wavelength on the
ring circumference) correspond to the displacements of the ring as
a whole in two orthogonal directions. The other (with $i=0$) is a
pure tangential mode describing the rotation of the ring as a
whole. The $k=0$ finite-frequency mode, on the other hand, is a
pure normal, breathing, mode that ``borrowed'' its strength from
the parent stretching oscillations of the linear
chain.\\

Equations (\ref{l},\ref{c}) can be used to study the continuum
limit, when, keeping the same radius $R$, we increase the number
of atoms $N\rightarrow\infty$ and bond length $l\rightarrow 0$.
Then, evidently,
\begin{equation}\label{contring}
\begin{array}{l}
\delta l/l \rightarrow \left[\partial v/\partial y + w/R\right], \\
\delta \chi/l \rightarrow -\partial \left[ \partial w/\partial y -
v/R \right]/\partial y.
\end{array}
\end{equation}
The expressions in brackets in Eq.~(\ref{contring}) provide a
recipe for a transition from the continuum model of a chain to the
continuum model of a ring. The subsequent analysis of vibrations
is straightforward with the wavevector $k$ quantized as $k R=n$ (n
being an integer) yielding three zero-frequency modes and the
breathing mode as in the discrete case above.

It should be noted here that in reality the very values of the
unit cell length $a$ and elastic constants $K$, $\Kb$ can in fact
somewhat differ for the ring and the linear chain, and the
difference would be $N-$dependent as determined by the equilibrium
bond length in the ring.  In this sense, what is compared in
Figure \ref{ring1} is the structure of the spectra for the same
values of the essential parameters upon  isometric mapping of the
linear system onto a circular ring. In addition, the elastic
energy of the ring could in general contain terms absent in the
energy of the chain, such as $\dcn \dln$. They are not deducible
from the functional form of the linear system and would have to be
explored on their own. We also note that Eq.~(\ref{u1}) can be
generalized to include longer range interactions.

\section{\label{newbz}Transverse quantization and unit cells}

\begin{figure}
\includegraphics[scale=0.5]{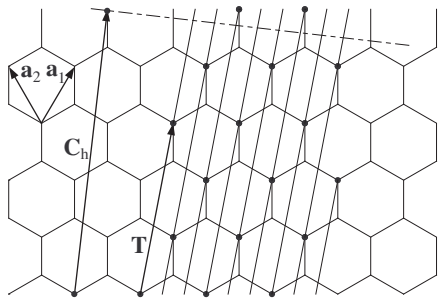}
\caption{\label{g3_2}Unwrapped (3,2) tube with carbons rearranged
to form a horizontal strip here. The tube axis is shown by the
dash-dotted line. Here $\dc=1$ and the $\TT$ vector,
Eq.~(\ref{TT}), is defined by the $(P,Q)$ pair (2,1). Small
circles connected by thinner lines indicate a sequence of carbons
reached by the $\TT-$translation. It is transparent that all
hexagons of the strip would be visited this way, forming an
effectively 1-$d$ enumeration of Eq.~(\ref{Ri}). The conventional
vector $\TTD=7\aone-8\atwo$ and such a unit cell would have 38
hexagons.}
\end{figure}

Two carbons connected by $\CC$ (Eq.~(\ref{CC})) on the graphene
plane correspond to the same carbon on the nanotube after
wrapping. The effective 2-$d$ cyclic condition can therefore be
written as
\begin{equation}\label{Rnm}
\Rnm + \CC = \Rnm, \ \ \ \Rnm= n\aone + m\atwo,
\end{equation}
$\Rnm$ being the position of one of the carbons (there are two of
them) of an arbitrary unit cell of graphene. If $\ko$ is the 2-$d$
wave vector of the band excitations, then in the infinite graphene
plane it would have two independent continuous components. In
nanotubes, the cyclic condition leads to the transverse
quantization of  Eq.~(\ref{quant}). In other words, wave vectors
$\ko=\kperp+\kparl$ allowed by Eq.~(\ref{quant}) lie only on
certain quantization lines in the reciprocal plane of graphene,
$\kperp$ being quantized according to Eq.~(\ref{quant}) and
$\kparl$ being actually a 1-$d$ continuous wave vector parallel to
the tube axis.

Let us define vector
\begin{equation}\label{TT}
\TT=P\aone+Q\atwo
\end{equation}
with integer $P$ and $Q$. From Eqs.~(\ref{CC}) and (\ref{TT}), one
finds
\begin{equation}\label{a12}
\aone=(-Q\CC+M\TT)/\Delta, \ \ \ \atwo=(P\CC-N\TT)/\Delta,
\end{equation}
where
\begin{equation}\label{determ}
\Delta=MP-NQ.
\end{equation}
Evidently, \textit{if} one can find such $P$ and $Q$ (of course,
we are interested in the ``smallest'' $P$ and $Q$) for a given
tube $(N,M)$ that $\Delta=\pm 1$ in Eq.~(\ref{determ}), then
original primitive vectors $\aone$ and $\atwo$ in Eq.~(\ref{a12})
will be represented through integer amounts of $\CC$ and $\TT$.
Correspondingly, an arbitrary vector $\Rnm$, Eq.~(\ref{Rnm}), will
be expressed through integer quantities of $\CC$ and $\TT$ as
well. Since the cyclic condition defines $\Rnm$ in Eq.~(\ref{Rnm})
with accuracy to $\CC$, this would actually mean that the position
of any hexagon in the unwrapped tube is determined through a
single vector $\TT$:
\begin{equation}\label{Ri}
\Rnm \rightarrow \Ri=i\TT,
\end{equation}
where, e.g., $i=nM-mN$ for $\Delta=1$. One can think of the
corresponding unit cell built of $\CC$ and $\TT$ that would
contain only 2 carbons in the cell. Of course, vector $\TT$
\textit{does not have to be parallel} to the tube axis. The
projection of $\TT$ on the nanotube axis $\TTparl=\sq a^{2}
\Delta/2\Ch$ -- it would be directed in opposite ways for
$\Delta=1$ vs $\Delta=-1$ -- and its modulus determines the
corresponding longitudinal period
\begin{equation}\label{aper}
\aper=\sq a^{2}/2\Ch,
\end{equation}
where $\Ch=|\CC|$. Note that vector $\TT$ is different from the
symmetry vector defined in Ref.~\onlinecite{book1}.

It is easy to see that the picture described in the previous
paragraph is indeed realized whenever $\dc=1$ -- we will call it
the irreducible case -- where $\dc$ is the greatest common divisor
of $N$ and $M$. (Considering only positive $N$ and $M$  does not
restrict the generality.) Then, in fact, any integer value of
$\Delta$ in Eq.~(\ref{determ}) can be established with an
appropriate choice of $P$ and $Q$. Examples of positive $(P,Q)$
pairs satisfying $\Delta=1$ are listed here as $(N,M) \rightarrow
(P,Q)$: $(3,1) \rightarrow (1,0)$, $(3,2) \rightarrow (2,1)$,
$(5,2) \rightarrow (3,1)$, and $(10,9) \rightarrow (9,8)$. An
illustration for the (3,2) tube is shown in Figure \ref{g3_2}. A
visual picture of the irreducible case is that of a 1-$d$ chain
(of period $|\TT|$) that is wrapped around the nanotube cylinder
with an appropriate helix angle, as is clearly seen from Figure
\ref{g3_2}.

\begin{figure}
\includegraphics[scale=0.45]{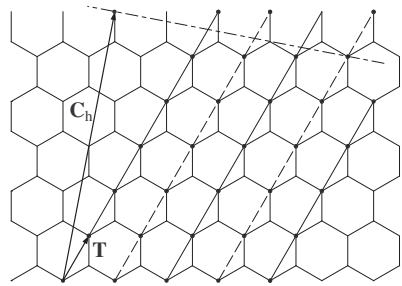}
\caption{\label{g4_2}Unwrapped (4,2) tube with carbons rearranged
to form a horizontal strip here. The tube axis is shown by the
dash-dotted line. Here $\dc=2$ and the $\TT$ vector,
Eq.~(\ref{TT}), is defined by the ``reduced'' $(P,Q)$ pair (1,0).
Two sequences of carbons resulting from $\TT-$translations are
shown by thinner lines: small circles of one sequence are
connected by the solid line, and circles of the other by the
dashed line.  Translation from one sequence to the other is
achieved with vector $\CC/2$. All hexagons of the strip would be
visited this way, forming effective enumeration of
Eq.~(\ref{Rij}). The conventional vector $\TTD=4\aone-5\atwo$ and
such a unit cell would have 28 hexagons.}
\end{figure}

``Reducible'' cases -- with $\dc > 1$ -- can be described in a
similar but somewhat more involved fashion. One would extract the
irreducible structure factors $\No$ and $\Mo$:
$$
N=\dc \No, \ \ \ M=\dc \Mo,
$$
so that the greatest common divisor of $\No$ and $\Mo$ equals 1.
Then the procedure described above for the irreducible case can be
applied for the geometry $(\No,\Mo)$ resulting in the first
translational vector $\TT$. In other words, integers $P$ and $Q$
of Eq.~(\ref{TT}) should satisfy condition
\begin{equation}\label{PQ1}
MP-NQ=\dc.
\end{equation}
 One however could not visit all hexagons by using only so
defined $\TT$. The needed second translational vector can be found
as $\CC/\dc$. The resulting enumeration of all hexagons will then
read as
\begin{equation}\label{Rij}
\Rnm \rightarrow \Rij=i\TT + j\CC/\dc, \ \ \ j=0,1,\ldots \dc-1.
\end{equation}
Index $i$ here is, as in Eq.~(\ref{Ri}), an arbitrary integer that
would define unique carbons while $j$ results only in $\dc$ unique
translations due to the cyclic condition in Eq.~(\ref{Rnm}). An
illustration for the (4,2) tube is shown in Figure \ref{g4_2}. A
visual picture of the irreducible case is then of $\dc$ 1-$d$
chains (of period $|\TT|$) that are wrapped around the nanotube
cylinder with an appropriate helix angle. The elementary unit cell
contains two carbons and built of vectors $\TT$ and $\CC/\dc$.
Longitudinal periodicity will again be determined by the
projection of vector $\TT$ on the nanotube axis which is
Eq.~(\ref{aper1}) becoming Eq.~(\ref{aper}) at $\dc=1$.

\begin{figure}
\includegraphics[scale=0.7]{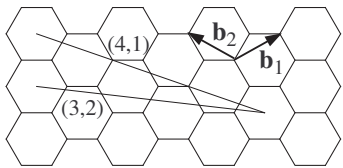}
\caption{\label{recipr1}An illustration of the range of variation
of $\kparl$ for (3,2) and (4,1) tubes. In the former case
$\rKt=2\bone-3\btwo$, in the latter case $\rKt=\bone-4\btwo$. The
corresponding ranges are shown as thick lines between the centers
of hexagons of the reciprocal lattice of graphene. Each line
crosses several hexagons. If all inside-a-hexagon segments are
displaced to one hexagon with a graphene reciprocal vector, they
would form a traditional picture of quantization lines in the
first BZ of graphene.}
\end{figure}

For the reciprocal lattice vectors, defined through
$\rKo\cdot\CC=2\pi$, $\rKt\cdot\TT=2\pi$, $\rKo\cdot\TT=0$, and
$\rKt\cdot\CC=0$, one readily obtains
$$
\rKo=-Q\bone+P\btwo, \ \ \rKt=\Mo\bone-\No\btwo,
$$
where $\bone$ and $\btwo$ are primitive graphene reciprocal
vectors defined with respect to $\aone$ and $\atwo$. One easily
finds that $|\rKt|$ indeed coincides with $2\pi/\aper$. An
illustration for cases of (3,2) and (4,1) tubes is given in Figure
\ref{recipr1}.

This way one arrives at the band excitation picture with 2$\dc$
(for each component of the wave function or polarization)
continuous bands corresponding to 2 atoms in the unit cell and
$\dc$ transverse quantization levels, e.g. with quantum numbers
\begin{equation}\label{ls}
l = -\dc/2+1, \ldots, 0, \ldots \dc/2.
\end{equation}
The latter define the quantized component of the parent 2-$d$
quasi-momentum $\ko$ perpendicular to the tube axis:
\begin{equation}\label{kperp}
\kpe=(2\pi/\Ch) l.
\end{equation}
The continuous component $\kp$ parallel to the tube axis, on the
other hand, would be defined within a BZ whose width is
\begin{equation}\label{kparl}
2\pi/\aper=4\pi \Ch/\sq \dc a^{2}.
\end{equation}
One can easily calculate that the resulted number of continuous
bands and the width of the BZ  lead to the correct total number of
states in the system, which corresponds to $2\Ch/\sq a^{2}$
hexagons per unit length of the nanotube.\\

The conclusion that there are only $\dc$ ``truly unique''
quantization levels can be confirmed another way as well. If $
\GG=\no\bone+\nt\btwo $ is the vector of the reciprocal lattice of
graphene, then $\kt=\ko + \GG$ is physically equivalent to $\ko$.
In general, $\ko$ and $\kt$  may correspond to different quantum
numbers in Eq.~(\ref{quant}), say $l$ and $l_{1}$.  One derives
the integer difference of these quantum numbers as
\begin{equation}\label{dl}
\dl=l-l_{1}=\no N+\nt M=\dc (\no\No+\nt\Mo).
\end{equation}

Once again, the last, ``irreducible'', factor in Eq.~(\ref{dl})
can take any integer values:
$$
\no N_{1}+\nt M_{1}=0, \pm 1, \pm 2, \ldots,
$$
with appropriate choices for integer $\no$ and $\nt$. It is then
clear that the number of unique sub-bands in the extended scheme
equals precisely $\dc$ -- one can, e.g., choose quantum numbers of
Eq.~(\ref{ls}) for ``independent'' sub-band indexing, all other
values of $l$ would be reducible to independent values with the
appropriate
adjustment for the one-dimensional continuous quasi-momentum $\kp$.\\

It is worth mentioning that in the $(N,M)$ family of nanotubes
with $N > M$, it is the armchair and zigzag nanotubes that have
maximal number of transverse quantization levels $\dc=N$. The
tubes immediately next to them, $(N,N-1)$ and $(N,1)$, on the
other hand, would have only one quantization level. The armchair
tubes have then the longitudinal period of $a/2$, twice as small
as is with the conventional definition.\cite{book1} One can easily
see it with $\TT$ vector defined by the
$(P,Q)$ pair (1,0).\\

\bibliography{../../pebib}

\begin{thebibliography}{26}
\expandafter\ifx\csname natexlab\endcsname\relax\def\natexlab#1{#1}\fi
\expandafter\ifx\csname bibnamefont\endcsname\relax
  \def\bibnamefont#1{#1}\fi
\expandafter\ifx\csname bibfnamefont\endcsname\relax
  \def\bibfnamefont#1{#1}\fi
\expandafter\ifx\csname citenamefont\endcsname\relax
  \def\citenamefont#1{#1}\fi
\expandafter\ifx\csname url\endcsname\relax
  \def\url#1{\texttt{#1}}\fi
\expandafter\ifx\csname urlprefix\endcsname\relax\def\urlprefix{URL }\fi
\providecommand{\bibinfo}[2]{#2}
\providecommand{\eprint}[2][]{\url{#2}}

\bibitem[{\citenamefont{Saito et~al.}(1998)\citenamefont{Saito, Dresselhaus,
  and Dresselhaus}}]{book1}
\bibinfo{author}{\bibfnamefont{R.}~\bibnamefont{Saito}},
  \bibinfo{author}{\bibfnamefont{G.}~\bibnamefont{Dresselhaus}},
  \bibnamefont{and} \bibinfo{author}{\bibfnamefont{M.~S.}
  \bibnamefont{Dresselhaus}}, \emph{\bibinfo{title}{Physical Properties of
  Carbon Nanotubes}} (\bibinfo{publisher}{Imperial College Press},
  \bibinfo{address}{London}, \bibinfo{year}{1998}).

\bibitem[{\citenamefont{Dresselhaus and Eklund}(2000)}]{phononreview}
\bibinfo{author}{\bibfnamefont{M.~S.} \bibnamefont{Dresselhaus}}
  \bibnamefont{and} \bibinfo{author}{\bibfnamefont{P.}~\bibnamefont{Eklund}},
  \bibinfo{journal}{Adv. Phys.} \textbf{\bibinfo{volume}{49}},
  \bibinfo{pages}{705} (\bibinfo{year}{2000}).

\bibitem[{\citenamefont{Cao et~al.}(2003)\citenamefont{Cao, Yan, Xiao, Tang,
  and Ding}}]{cao}
\bibinfo{author}{\bibfnamefont{J.~X.} \bibnamefont{Cao}},
  \bibinfo{author}{\bibfnamefont{X.~H.} \bibnamefont{Yan}},
  \bibinfo{author}{\bibfnamefont{Y.}~\bibnamefont{Xiao}},
  \bibinfo{author}{\bibfnamefont{Y.}~\bibnamefont{Tang}}, \bibnamefont{and}
  \bibinfo{author}{\bibfnamefont{J.~W.} \bibnamefont{Ding}},
  \bibinfo{journal}{Phys. Rev. B} \textbf{\bibinfo{volume}{67}},
  \bibinfo{pages}{045413} (\bibinfo{year}{2003}).

\bibitem[{\citenamefont{S\'{a}nchez-Portal
  et~al.}(1999)\citenamefont{S\'{a}nchez-Portal, Artacho, Soler, Rubio, and
  Ordej\'{o}n}}]{abini1}
\bibinfo{author}{\bibfnamefont{D.}~\bibnamefont{S\'{a}nchez-Portal}},
  \bibinfo{author}{\bibfnamefont{E.}~\bibnamefont{Artacho}},
  \bibinfo{author}{\bibfnamefont{J.~M.} \bibnamefont{Soler}},
  \bibinfo{author}{\bibfnamefont{A.}~\bibnamefont{Rubio}}, \bibnamefont{and}
  \bibinfo{author}{\bibfnamefont{P.}~\bibnamefont{Ordej\'{o}n}},
  \bibinfo{journal}{Phys. Rev. B} \textbf{\bibinfo{volume}{59}},
  \bibinfo{pages}{12678} (\bibinfo{year}{1999}).

\bibitem[{\citenamefont{Yu et~al.}(1995)\citenamefont{Yu, Kalia, and
  Vashishta}}]{tbmd}
\bibinfo{author}{\bibfnamefont{J.}~\bibnamefont{Yu}},
  \bibinfo{author}{\bibfnamefont{R.~K.} \bibnamefont{Kalia}}, \bibnamefont{and}
  \bibinfo{author}{\bibfnamefont{P.}~\bibnamefont{Vashishta}},
  \bibinfo{journal}{J. Chem. Phys.} \textbf{\bibinfo{volume}{103}},
  \bibinfo{pages}{6697} (\bibinfo{year}{1995}).

\bibitem[{\citenamefont{Nishiguchi et~al.}(1997)\citenamefont{Nishiguchi, Ando,
  and Wybourne}}]{nishi1}
\bibinfo{author}{\bibfnamefont{N.}~\bibnamefont{Nishiguchi}},
  \bibinfo{author}{\bibfnamefont{Y.}~\bibnamefont{Ando}}, \bibnamefont{and}
  \bibinfo{author}{\bibfnamefont{M.~N.} \bibnamefont{Wybourne}},
  \bibinfo{journal}{J. Phys.: Condens. Matter} \textbf{\bibinfo{volume}{9}},
  \bibinfo{pages}{5751} (\bibinfo{year}{1997}).

\bibitem[{\citenamefont{Rego and Kirczenow}(1998)}]{rego1}
\bibinfo{author}{\bibfnamefont{L.~G.~C.} \bibnamefont{Rego}} \bibnamefont{and}
  \bibinfo{author}{\bibfnamefont{G.}~\bibnamefont{Kirczenow}},
  \bibinfo{journal}{Phys. Rev. Lett.} \textbf{\bibinfo{volume}{81}},
  \bibinfo{pages}{232} (\bibinfo{year}{1998}).

\bibitem[{\citenamefont{Schwab et~al.}(2000)\citenamefont{Schwab, Henriksen,
  Worlock, and Roukes}}]{expquant}
\bibinfo{author}{\bibfnamefont{K.}~\bibnamefont{Schwab}},
  \bibinfo{author}{\bibfnamefont{E.~A.} \bibnamefont{Henriksen}},
  \bibinfo{author}{\bibfnamefont{J.~M.} \bibnamefont{Worlock}},
  \bibnamefont{and} \bibinfo{author}{\bibfnamefont{M.~L.}
  \bibnamefont{Roukes}}, \bibinfo{journal}{Nature}
  \textbf{\bibinfo{volume}{404}}, \bibinfo{pages}{974} (\bibinfo{year}{2000}).

\bibitem[{\citenamefont{Eliel et~al.}(2001)\citenamefont{Eliel, Wilen, and
  Doyle}}]{stereo1}
\bibinfo{author}{\bibfnamefont{E.~L.} \bibnamefont{Eliel}},
  \bibinfo{author}{\bibfnamefont{S.~H.} \bibnamefont{Wilen}}, \bibnamefont{and}
  \bibinfo{author}{\bibfnamefont{M.~P.} \bibnamefont{Doyle}},
  \emph{\bibinfo{title}{Basic Organic Stereochemistry}}
  (\bibinfo{publisher}{Wiley}, \bibinfo{address}{New York},
  \bibinfo{year}{2001}).

\bibitem[{\citenamefont{Yu and Cardona}(1999)}]{cardona}
\bibinfo{author}{\bibfnamefont{P.~Y.} \bibnamefont{Yu}} \bibnamefont{and}
  \bibinfo{author}{\bibfnamefont{M.}~\bibnamefont{Cardona}},
  \emph{\bibinfo{title}{Fundamentals of Semiconductors}}
  (\bibinfo{publisher}{Springer}, \bibinfo{address}{Berlin},
  \bibinfo{year}{1999}).

\bibitem[{\citenamefont{\v{S}. Marku\v{s}}(1988)}]{bookmarkus}
\bibinfo{author}{\bibnamefont{\v{S}. Marku\v{s}}}, \emph{\bibinfo{title}{The
  Mechanics of Vibrations of Cylindrical Shells}}
  (\bibinfo{publisher}{Elsevier}, \bibinfo{address}{Amsterdam},
  \bibinfo{year}{1988}).

\bibitem[{\citenamefont{Landau and Lifshitz}(1986)}]{ll8}
\bibinfo{author}{\bibfnamefont{L.~D.} \bibnamefont{Landau}} \bibnamefont{and}
  \bibinfo{author}{\bibfnamefont{E.~M.} \bibnamefont{Lifshitz}},
  \emph{\bibinfo{title}{Elasticity Theory}} (\bibinfo{publisher}{Pergamon},
  \bibinfo{address}{Oxford}, \bibinfo{year}{1986}).

\bibitem[{\citenamefont{Sirenko
  et~al.}(1996{\natexlab{a}})\citenamefont{Sirenko, Stroscio, and
  Kim}}]{sirenko1}
\bibinfo{author}{\bibfnamefont{Y.~M.} \bibnamefont{Sirenko}},
  \bibinfo{author}{\bibfnamefont{M.~A.} \bibnamefont{Stroscio}},
  \bibnamefont{and} \bibinfo{author}{\bibfnamefont{K.~W.} \bibnamefont{Kim}},
  \bibinfo{journal}{Phys. Rev. E} \textbf{\bibinfo{volume}{53}},
  \bibinfo{pages}{1003} (\bibinfo{year}{1996}{\natexlab{a}}).

\bibitem[{\citenamefont{Sirenko
  et~al.}(1996{\natexlab{b}})\citenamefont{Sirenko, Stroscio, and
  Kim}}]{sirenko2}
\bibinfo{author}{\bibfnamefont{Y.~M.} \bibnamefont{Sirenko}},
  \bibinfo{author}{\bibfnamefont{M.~A.} \bibnamefont{Stroscio}},
  \bibnamefont{and} \bibinfo{author}{\bibfnamefont{K.~W.} \bibnamefont{Kim}},
  \bibinfo{journal}{Phys. Rev. E} \textbf{\bibinfo{volume}{54}},
  \bibinfo{pages}{1816} (\bibinfo{year}{1996}{\natexlab{b}}).

\bibitem[{\citenamefont{Stroscio and Dutta}(2001)}]{bookphonons}
\bibinfo{author}{\bibfnamefont{M.~A.} \bibnamefont{Stroscio}} \bibnamefont{and}
  \bibinfo{author}{\bibfnamefont{M.}~\bibnamefont{Dutta}},
  \emph{\bibinfo{title}{Phonons in Nanostructures}}
  (\bibinfo{publisher}{Cambridge University Press},
  \bibinfo{address}{Cambridge}, \bibinfo{year}{2001}).

\bibitem[{\citenamefont{Madelung}(1978)}]{madelung}
\bibinfo{author}{\bibfnamefont{O.}~\bibnamefont{Madelung}},
  \emph{\bibinfo{title}{Introduction to Solid-State Theory}}
  (\bibinfo{publisher}{Springer}, \bibinfo{address}{Berlin},
  \bibinfo{year}{1978}).

\bibitem[{\citenamefont{Samsonidze et~al.}(2003)\citenamefont{Samsonidze,
  Saito, Jorio, \mbox{Souza Filho}, Gr\"{u}neis, Pimenta, Dresselhaus, and
  Dresselhaus}}]{newforce}
\bibinfo{author}{\bibfnamefont{G.~G.} \bibnamefont{Samsonidze}},
  \bibinfo{author}{\bibfnamefont{R.}~\bibnamefont{Saito}},
  \bibinfo{author}{\bibfnamefont{A.}~\bibnamefont{Jorio}},
  \bibinfo{author}{\bibfnamefont{A.~G.} \bibnamefont{\mbox{Souza Filho}}},
  \bibinfo{author}{\bibfnamefont{A.}~\bibnamefont{Gr\"{u}neis}},
  \bibinfo{author}{\bibfnamefont{M.~A.} \bibnamefont{Pimenta}},
  \bibinfo{author}{\bibfnamefont{G.}~\bibnamefont{Dresselhaus}},
  \bibnamefont{and} \bibinfo{author}{\bibfnamefont{M.~S.}
  \bibnamefont{Dresselhaus}}, \bibinfo{journal}{Phys. Rev. Lett.}
  \textbf{\bibinfo{volume}{90}}, \bibinfo{pages}{027403}
  (\bibinfo{year}{2003}).

\bibitem[{\citenamefont{Love}(1944)}]{rayleigh}
\bibinfo{author}{\bibfnamefont{A.~E.~H.} \bibnamefont{Love}},
  \emph{\bibinfo{title}{Treatise on the Mathematical Theory of Elasticity}}
  (\bibinfo{publisher}{Dover}, \bibinfo{address}{New York},
  \bibinfo{year}{1944}).

\bibitem[{\citenamefont{Rayleigh}(1976)}]{love}
\bibinfo{author}{\bibfnamefont{J.~W.~S.} \bibnamefont{Rayleigh}},
  \emph{\bibinfo{title}{The theory of sound}} (\bibinfo{publisher}{Dover},
  \bibinfo{address}{New York}, \bibinfo{year}{1976}).

\bibitem[{\citenamefont{Timoshenko and Woinowsky-Krieger}(1959)}]{timoshenko}
\bibinfo{author}{\bibfnamefont{S.~P.} \bibnamefont{Timoshenko}}
  \bibnamefont{and}
  \bibinfo{author}{\bibfnamefont{S.}~\bibnamefont{Woinowsky-Krieger}},
  \emph{\bibinfo{title}{Theory of Plates and Shells}}
  (\bibinfo{publisher}{McGraw-Hill}, \bibinfo{address}{New York},
  \bibinfo{year}{1959}).

\bibitem[{\citenamefont{Timoshenko and Gere}(1961)}]{timoshenko1}
\bibinfo{author}{\bibfnamefont{S.~P.} \bibnamefont{Timoshenko}}
  \bibnamefont{and} \bibinfo{author}{\bibfnamefont{J.~M.} \bibnamefont{Gere}},
  \emph{\bibinfo{title}{Theory of Elastic Stability}}
  (\bibinfo{publisher}{McGraw-Hill}, \bibinfo{address}{New York},
  \bibinfo{year}{1961}).

\bibitem[{\citenamefont{Arnold and Warburton}(1949)}]{arnold}
\bibinfo{author}{\bibfnamefont{R.~N.} \bibnamefont{Arnold}} \bibnamefont{and}
  \bibinfo{author}{\bibfnamefont{G.~B.} \bibnamefont{Warburton}},
  \bibinfo{journal}{Proc. Roy. Soc. A} \textbf{\bibinfo{volume}{197}},
  \bibinfo{pages}{238} (\bibinfo{year}{1949}).

\bibitem[{\citenamefont{Yakobson et~al.}(1996)\citenamefont{Yakobson, Brabec,
  and Bernholc}}]{yakobson}
\bibinfo{author}{\bibfnamefont{B.~I.} \bibnamefont{Yakobson}},
  \bibinfo{author}{\bibfnamefont{C.~J.} \bibnamefont{Brabec}},
  \bibnamefont{and} \bibinfo{author}{\bibfnamefont{J.}~\bibnamefont{Bernholc}},
  \bibinfo{journal}{Phys. Rev. Lett.} \textbf{\bibinfo{volume}{76}},
  \bibinfo{pages}{2511} (\bibinfo{year}{1996}).

\bibitem[{\citenamefont{K\"{u}rti et~al.}(2003)\citenamefont{K\"{u}rti,
  Z\'{o}lyomi, Kertesz, and Sun}}]{rbmmiklosh}
\bibinfo{author}{\bibfnamefont{J.}~\bibnamefont{K\"{u}rti}},
  \bibinfo{author}{\bibfnamefont{V.}~\bibnamefont{Z\'{o}lyomi}},
  \bibinfo{author}{\bibfnamefont{M.}~\bibnamefont{Kertesz}}, \bibnamefont{and}
  \bibinfo{author}{\bibfnamefont{G.}~\bibnamefont{Sun}}, \bibinfo{journal}{New
  Journal of Physics} \textbf{\bibinfo{volume}{5}}, \bibinfo{pages}{125.1}
  (\bibinfo{year}{2003}).

\bibitem[{\citenamefont{Gartstein et~al.}(2003)\citenamefont{Gartstein,
  Zakhidov, and Baughman}}]{gzb2}
\bibinfo{author}{\bibfnamefont{Y.~N.} \bibnamefont{Gartstein}},
  \bibinfo{author}{\bibfnamefont{A.~A.} \bibnamefont{Zakhidov}},
  \bibnamefont{and} \bibinfo{author}{\bibfnamefont{R.~H.}
  \bibnamefont{Baughman}}, \bibinfo{journal}{Phys. Rev. B}
  \textbf{\bibinfo{volume}{68}}, \bibinfo{pages}{115415}
  (\bibinfo{year}{2003}).

\bibitem[{\citenamefont{Robertson et~al.}(1992)\citenamefont{Robertson,
  Brenner, and Mintmire}}]{mint1}
\bibinfo{author}{\bibfnamefont{D.~H.} \bibnamefont{Robertson}},
  \bibinfo{author}{\bibfnamefont{D.~W.} \bibnamefont{Brenner}},
  \bibnamefont{and} \bibinfo{author}{\bibfnamefont{J.~W.}
  \bibnamefont{Mintmire}}, \bibinfo{journal}{Phys. Rev. B}
  \textbf{\bibinfo{volume}{45}}, \bibinfo{pages}{12592} (\bibinfo{year}{1992}).

\end{thebibliography}

\end{document}